\documentclass[12pt,amslatex]{article}
\usepackage{amsmath,amsfonts,amssymb}

\newlength{\dinwidth}
\newlength{\dinmargin}
\newtheorem{prop}{Proposition}[section]

\newtheorem{definition}{Definition}[section]
\newenvironment{proof}{\medskip \noindent
            {\bf Proof.}}{ \hfill $\square$ \medskip}
\def\cA{{\cal A}}
\def\cB{{\cal B}}
\def\cC{{\cal C}}

\def\cH{{\cal H}}

\def\cK{{\cal K}}

\def\cM{{\cal M}}
\def\cN{{\cal N}}
\def\cP{{\cal P}}

\def\cS{{\cal S}}

\def\cZ{{\cal Z}}
\def\idty{{\leavevmode\hbox{\rm 1\kern -.3em I}}}

\def\As{{\cal A}}
\def\Bs{{\cal B}}
\def\Cs{{\cal C}}
\def\Ds{{\cal D}}

\def\Fs{{\cal F}}

\def\Hs{{\cal H}}

\def\Ks{{\cal K}}
\def\Ls{{\cal L}}
\def\Ms{{\cal M}}
\def\Ns{{\cal N}}
\def\Os{{\cal O}}
\def\Ps{{\cal P}}

\def\Rs{{\cal R}}
\def\Ss{{\cal S}}
\def\Ts{{\cal T}}

\def\Ws{{\cal W}}

\def\Zs{{\cal Z}}

\def\Pid{{\Ps_+ ^{\uparrow}}}

\def\idty{{\leavevmode\hbox{\rm 1\kern -.3em I}}}
\def\RR{{\mathbb R}}
\def\IN{{\mathbb N}}

\def\CC{{\mathbb C}}

\def\Pid{{\Ps_+ ^{\uparrow}}}

\newcommand{\ie}{{\it i.e.\ }}
\newcommand{\eg}{{\it e.g.\ }}

\def\tr{{\rm Tr}}

\def\C{$C^{\ast}$-}
\def\W{$W^{\ast}$-}
\def\bbbr{{\rm I\!R}} 

\def\PH{{\cal P} ({\cal H})}
\def\BH{{\cal B} ({\cal H} )}

\def\PM{{\cal P} ({\cal M})}
\def\c{{\bot}} 

\def\vagy{\vee}
\def\es{\wedge}

\def\sup{{\rm sup}}

\begin{document}

\title{Quantum Probability Theory}

\author{{\Large Mikl\'os R\'edei\,$^{a,}$\thanks{Work supported by the 
Hungarian Scientific Research Fund (OTKA); contract number: T 043642.} \ and 
\
Stephen J.\ Summers\,$^b$ }\\[5mm]
${}^a$ Department of History and Philosophy of Science, \\
Lor\'and E\"otv\"os University, \\
P.O. Box 32, H-1518 Budapest 112, Hungary  \\[2mm]
${}^b$ Department of Mathematics,
University of Florida, \\ Gainesville FL 32611, USA}

\date{July 27, 2006}
\maketitle

\abstract{\noindent The mathematics of classical probability theory
was subsumed into classical measure theory by Kolmogorov in
1933. Quantum theory as nonclassical probability theory was
incorporated into the beginnings of noncommutative measure theory by
von Neumann in the early thirties, as well. To precisely this end, von
Neumann initiated the study of what are now called von Neumann
algebras and, with Murray, made a first classification of such
algebras into three types. The nonrelativistic quantum theory of
systems with finitely many degrees of freedom deals exclusively with
type I algebras. However, for the description of further quantum
systems, the other types of von Neumann algebras are indispensable.
The paper reviews quantum probability theory in terms of general von
Neumann algebras, stressing the similarity of the conceptual structure
of classical and noncommutative probability theories and emphasizing
the correspondence between the classical and quantum concepts, though
also indicating the nonclassical nature of quantum probabilistic
predictions. In addition, differences between the probability theories
in the type I, II and III settings are explained.  A brief description
is given of quantum systems for which probability theory based on type
I algebras is known to be insufficient. These illustrate the physical
significance of the previously mentioned differences.}

\newpage
\noindent
{\bf Contents}\\

\noindent
\ref{intro}. Introduction\\
\ref{algebra}. Algebras of Bounded Operators\\
\ref{classical}. Classical Probability Theory\\
\ref{quantum}. Noncommutative Probability Theory\\
\ref{QM}. Quantum Mechanics: Type I Noncommutative Probability Theory\\
\ref{necessity}. The Necessity of Non-Type--I Probability Spaces in
Physics\\
\indent \ref{qstatphys}. Quantum Statistical Mechanics\\
\indent \ref{brief}. Brief Return to General Quantum Statistical Mechanics\\
\indent \ref{QFT}. Local Relativistic Quantum Field Theory\\
\ref{differences}. Some Differences of Note\\
\indent \ref{Bell}. Entanglement and Bell's Inequalities\\
\indent \ref{independence}. Independence\\
\indent \ref{conditional}. Conditional Expectations \\
\indent \ref{contrast}. Further Comments on Type I versus Type III\\
\ref{closing}. Closing Words

\bigskip

\section{Introduction\label{intro}}

     In 1933, probability theory found its modern form in the classic
work of N.S. Kolmogorov \cite{Kol}, where it was treated axiomatically
as a branch of classical measure theory. There was, of course,
significant prior work in that direction --- in his own (translated) words
\cite{Kol}, ``While a conception of probability theory based on the
above general viewpoints has been current for some time among certain
mathematicians, there was lacking a complete exposition of the whole
system, free of extraneous complications.'' The historical development
leading to Kolmogorov's work was lengthy and involved mathematical,
physical and philosophical considerations (some stages of this
evolution are discussed in \cite{Tod,Plato}). A notable event in this
development was D. Hilbert's famous 1900 lecture in Paris on open
problems in mathematics, wherein he called for an axiomatic treatment
of probability theory (Hilbert's Sixth Problem) \cite{Hilbert6}.

     In the same problem, Hilbert also called for an axiomatization of
physics, and he himself made important contributions to the subject
(cf. \cite{Wight} for an overview). In the winter term of 1926--1927,
he gave a series of lectures on the newly emergent quantum mechanics,
which were prepared in collaboration with his assistants, L. Nordheim
and J. von Neumann. These were published in a joint paper \cite{HvN},
which was followed up by further papers of von Neumann
\cite{Neumann1927b,Neumann1927c,Neumann1927d}.  This approach
culminated in von Neumann's axiomatization of nonrelativistic quantum
mechanics in Hilbert space \cite{Neumann1932}.\footnote{To quote from
\cite{Wight}: ``I do not know whether Hilbert regarded von Neumann's
book as the fulfillment of the axiomatic method applied to quantum
mechanics, but, viewed from afar, that is the way it looks to me. In
fact, in my opinion, it is the most important axiomatization of a
physical theory up to this time.''}

     Motivated by this development, F.J. Murray and von Neumann
commenced a study of algebras of bounded operators on Hilbert space
\cite{MvN}. Over time, it slowly became clear that the same
mathematical tools were of direct relevance to the quantum theories of
more complicated systems, such as quantum statistical mechanics and
relativistic quantum field theory. As both the physical theories
and the mathematical ideas were refined and generalized, a probability
theory emerged which included both classical probability theory and
quantum theory as special cases.

     In this paper there will be no attempt to describe this
historical development. Rather, we shall present an overview of this
unification, as well as some of the similarities and differences which
this unification both relies on and reveals. We shall also clarify
some of the corresponding differences and similarities in the probability
theories appropriate for the treatment of quantum theories involving
only finitely many degrees of freedom and those involving infinitely
many degrees of freedom. Given page constraints, we shall
be obliged to treat these matters rather summarily, but further
references will be provided for the interested reader. Nothing will be
said about the branch of noncommutative probability theory known as
free probability theory, but see \cite{VDN}. Nor shall various
extensions of noncommutative probability theory to more general
algebraic or functional analytic structures be further mentioned.

     We address this paper to readers who have a familiarity with
elementary functional analysis (Hilbert spaces and Banach spaces),
probability theory and quantum mechanics, but who are not experts in
operator algebra theory, noncommutative probability theory or the
mathematical physics of quantum systems with infinitely many degrees
of freedom.

     We shall begin with a brief introduction to the mathematical
framework of operator algebra theory, within which this unification
has been accomplished.  In Section \ref{classical}, we describe
classical probability theory from the point of view of operator
algebra theory.  This should help the reader in Section \ref{quantum}
to recognize more readily the probability theory inherent in the
theory of normal states on von Neumann algebras, which is the setting
of noncommutative probability theory. Classical probability theory
finds its place therein as the special case where the von Neumann
algebra is abelian. Nonrelativistic quantum mechanics is then
understood in Section \ref{QM} as the special case where the von
Neumann algebra is a nonabelian type I algebra.

     Although the analogies and differences between classical
probability theory and nonrelativistic quantum mechanics are fairly
well known, the same cannot be said about the mathematical physics of
quantum systems with infinitely many degrees of freedom. We therefore
indicate the {\it necessity} of going beyond the type I case in
Section \ref{necessity}, where we discuss quantum statistical
mechanics and relativistic quantum field theory, showing how non-type--I
algebras arise in situations of immediate physical interest. Some of
the physically relevant differences between abelian (classical)
probability theory, nonabelian type I probability theory and non-type--I
probability theory will be indicated in Section \ref{differences}.

\section{Algebras of Bounded Operators} \label{algebra}

     In this section we shall briefly describe the aspects of
operator algebra theory which are most relevant to our topic.
For further details, see the texts \cite{Sak,Tak,KR}.

     Let $\Bs(\Hs)$ denote the set of all bounded operators on a
complex Hilbert space\footnote{For convenience, we shall always assume
our Hilbert spaces to be separable.} $\cH$. $\Bs(\Hs)$ is a complex
vector space under pointwise addition and scalar multiplication and a
complex algebra under the additional operation of composition. Adding
the involution $A \mapsto A^*$ of taking adjoints, $\Bs(\Hs)$ is a
*-algebra. Under the operator norm topology $\Bs(\Hs)$ is a Banach
space. A subalgebra $\cC \subset \Bs(\Hs)$ is called a (concrete)
{\it \C algebra} if it is closed with respect to the adjoint operation and
is closed in the operator norm topology.\footnote{There is a useful
notion of abstract \C algebra which does not refer to a Hilbert space:
an involutive Banach algebra $\cA$ which has the property
$\|AA^*\|=\|A\|\,\|A^*\|$ for every $A\in\cA$ is called an abstract \C
algebra. However, it is known that every abstract \C algebra is
isomorphic to a concrete \C algebra, so there is no loss of generality
to restrict our attention here to the concrete case and to drop the
qualifying adjective henceforth.}  The latter requirement means that
if $\{ A_n \}$ is a sequence of operators from $\cC$ which converges
in norm to some $A \in \Bs(\Hs)$, then $A\in\cC$. A \C algebra is
thus a Banach space with respect to this topology, since $\Bs(\Hs)$
is such a Banach space. We shall always
assume that the \C algebras $\Cs$ we employ are unital, \ie $I \in
\Cs$, where $I$ is the identity transformation on $\Hs$. Note that
$\Bs(\Hs)$ is itself a \C algebra.  Note further that if $X$ is a
compact Hausdorff space, then the algebra $C(X)$ of all continuous
complex-valued functions on $X$ supplied with the norm
$$ \Vert f \Vert = \sup\{ \vert f(x) \vert : x \in X \} $$
is an abelian \C algebra. It is noteworthy that every abelian
\C algebra is isomorphic to $C(X)$ for some (up to homeomorphism
unique) compact Hausdorff space $X$.

     Since every \C algebra $\cC$ is a Banach space, its topological
dual $\Cs^*$, consisting of continuous linear maps from $\Cs$ into the
complex numbers $\CC$, is also a Banach space. A {\it state} $\phi$ on a \C
algebra $\cC$ is such a map $\phi \in \Cs^*$ which is also positive,
\ie $\phi(A^* A) \geq 0$ for all $A \in \Cs$, and normalized, \ie
$\phi(I) = 1$. Given a state $\phi$ on $\cC$, one can construct a
Hilbert space $\cH_{\phi}$, a distinguished unit vector $\Omega_\phi
\in \cH_{\phi}$ and a \C algebra homomorphism $\pi_{\phi}\colon \cC
\to \cB(\cH_{\phi})$, so that $\pi_{\phi}(\cC)$ is a \C algebra acting
on the Hilbert space $\cH_{\phi}$, the set of vectors
$\pi_{\phi}(\cC)\Omega_\phi = \{ \pi_\phi(A)\Omega : A \in \Cs \}$ 
is dense in $\Hs_\phi$ and
$$\phi(A) = \langle \Omega_\phi, \pi_\phi(A) \Omega_\phi \rangle \, ,
\, A \in \Cs \, .$$
The triple $(\Hs_\phi,\Omega_\phi,\pi_\phi)$ is
uniquely determined up to unitary equivalence by these properties, and
$\pi_{\phi}$ is called the GNS representation of $\cC$ determined by
$\phi$.

     A {\it von Neumann algebra} $\Ms$ is a \C algebra which is also
closed in the strong operator topology. The latter requirement means
that if $\{ A_n \}$ is a sequence of operators from $\Ms$ such that
there exists an $A \in \Bs(\Hs)$ so that for all $\Phi\in\cH$
$A_n\Phi$ converges to $A\Phi$,  then $A\in\Ms$. In particular,
$\Bs(\Hs)$ is a  von Neumann algebra. The operator norm topology
is strictly stronger than the strong operator topology when $\Hs$ is
infinite dimensional. Hence, every von Neumann algebra is a \C algebra,
but the converse is false.  When $\Hs$ is finite dimensional, these
two topologies coincide and there is no distinction between \C \, and
von Neumann algebras.

     A remarkable fact is that a \C algebra $\Ms$ is a von Neumann algebra
if and only if there exists a Banach space $\Bs$ such that $\Ms$ is
(isomorphic to) the Banach dual of $\Bs$. If $\Bs$ exists, then it is unique
and is called the predual $\Ms_*$ of $\Ms$. From Banach space theory,
there is a canonical isometric embedding of $\Ms_*$ into $\Ms^*$
(using $\phi(A) = A(\phi)$, for all $\phi \in \Ms_*, A \in \Ms$). 
A {\it normal state} $\phi$ on
a von Neumann algebra $\Ms$ is a state which lies in (the embedded
image of) $\Ms_*$. Normal states are characterized by an
additional continuity property:
$\phi(\sup_\alpha A_\alpha) = \sup_\alpha \phi(A_\alpha)$,
for any uniformly bounded increasing net $\{ A_\alpha \}$ of positive
elements of $\Ms$. This continuity property is equivalent to the
following property:
\begin{equation}  \label{normal}
\phi(\sum_{n \in \IN} P_n) = \sum_{n \in \IN} \phi(P_n) \, ,
\end{equation}
for any countable family $\{ P_n \}_{n \in \IN}$ of mutually
orthogonal projections in $\Ms$. One can therefore define normal states on
$\Ms$ to be states satisfying (\ref{normal}).

     Every normal state on the
von Neumann algebra $\Bs(\Hs)$ is given by $\phi(A) = \tr(\rho A)$,
$A \in \Bs(\Hs)$, for some unique {\it density matrix} $\rho$ acting on
$\Hs$, \ie $0 \leq \rho = \rho^* \in \Bs(\Hs)$ such that
$\tr(\rho) = 1$. In other words, the predual of $\Bs(\Hs)$ is (isometrically
isomorphic to) the Banach space $\Ts(\Hs)$ of all trace-class operators on
$\Hs$ with the trace norm. A special case of such normal states is
constituted by  the vector states: if $\Phi \in \Hs$ is a unit
vector and $P_\Phi \in \Bs(\Hs)$ is the orthogonal projection onto the
one dimensional subspace of $\Hs$ spanned by $\Phi$, the corresponding
vector state is given by
$$\phi(A) = \langle \Phi, A \Phi \rangle = \tr(P_\Phi A) \, , \,
A \in \Bs(\Hs) \, .$$
It is a fairly straightforward application of the Hahn--Banach Theorem
to see that if $\Ms \subset \Bs(\Hs)$, every state on $\Ms$ is the
restriction to $\Ms$ of a state on $\Bs(\Hs)$. But it is much less
obvious, though equally true, that every normal state on $\Ms$ is the
restriction of a normal state on $\Bs(\Hs)$. Hence, for any normal
state $\phi \in \Ms^*$ there exists a (no longer necessarily unique)
density matrix $\rho \in \Bs(\Hs)$ such that
$\phi(A) = \tr(\rho A)$, for all $A \in \Ms$.

     A state $\phi$ on a \C algebra $\Cs$ is {\it faithful} if
$0 \leq A \in \Cs$ and $\phi(A) = 0$ entail $A = 0$, and $\phi$
is {\it tracial} if $\phi(AB) = \phi(BA)$, for all $A,B \in \Cs$.
If $\Hs$ has dimension $n$, then
$\widehat{\tr}(A) = \frac{1}{n} \tr(A)$, $A \in \Bs(\Hs)$,  is
called the {\it normalized trace} on $\Bs(\Hs)$ and is a faithful
normal tracial state. But if $\Hs$ is infinite dimensional, then there
exists no faithful normal tracial state on $\Bs(\Hs)$.

     Let $(X,\cS,\mu)$ be a finite measure space, where $X$ is a set,
$\cS$ is a $\sigma$-algebra of subsets of $X$ and $\mu$ is a finite
$\sigma$-additive measure on $\Ss$. Let $L^\infty(X,\Ss,\mu)$ denote
the Banach space  of all essentially bounded complex-valued functions
on $X$ supplied with the (essential) supremum norm, and let $L^1(X,\Ss,\mu)$
be the Banach space of $\mu$-integrable complex-valued functions on
$X$. Then the dual of $L^1(X,\Ss,\mu)$ is $L^\infty(X,\Ss,\mu)$. For
our purposes, the latter is viewed as an algebra of multiplication
operators acting on the Hilbert space $L^2(X,\Ss,\mu)$ of all square
integrable (with respect to $\mu$) complex-valued functions on
$X$. That is to say
$$ (f g)(x) = f(x) \, g(x) \, , \, x \in X \, , \, f \in L^\infty(X,\Ss,\mu)
\, , \, g \in L^2(X,\Ss,\mu) $$
defines for each $f \in L^\infty(X,\Ss,\mu)$ a linear mapping
$M_f \in \Bs(L^2(X,\Ss,\mu))$ by $M_f(g) = fg$, $g \in L^2(X,\Ss,\mu)$.
Note that $\Vert M_f \Vert = \Vert f \Vert_\infty$, so that
$L^\infty(X,\Ss,\mu) \ni f \mapsto M_f \in \Bs(L^2(X,\Ss,\mu))$
is an isometry. Hence, $L^\infty(X,\Ss,\mu)$ is a von Neumann algebra
with predual $L^1(X,\Ss,\mu)$.
Conversely, every abelian von Neumann algebra is isomorphic to
$L^\infty(X,\Ss,\mu)$ for some (up to isomorphism unique) localizable
measure space (\ie a direct sum of finite measure spaces).  Moreover,
since von Neumann algebras are \C algebras, an abelian von Neumann
algebra $\Ms$ is also \C isomorphic to $C(Y)$, for some compact Hausdorff 
space
$Y$. The normal states on $\Ms$ are determined precisely by the Radon
probability measures on $Y$. 

     If $S$ is any subset of $\Bs(\Hs)$, then its commutant
$S'$ is the set of bounded operators which commute with every element
in $S$, \ie
$$ S'\equiv\{B \in \Bs(\Hs) : AB=BA, \mbox{\ for all\ } A\in S\} \, . $$
The operation of taking the commutant can be iterated, $S''\equiv
(S')'$, and it is clear that $S \subset S''$. Von Neumann's {\it double
commutant theorem} asserts if $S$ is a
subalgebra containing $I$ and closed under taking adjoints, then $S''$
is the closure of $S$ in the strong operator topology. The double
commutant theorem implies that a subset $S \subset \Bs(\Hs)$ is a von
Neumann algebra if and only if $S = S''$. This then is a purely
algebraic characterization of von Neumann algebras. It also follows
that $S'$ is a von Neumann algebra. A von Neumann algebra $\cM$ is
called a {\em factor} if the only elements in $\cM$ which commute with
every other element in $\cM$ are the constant multiples of the
identity, \ie if $\cM\cap\cM'= \CC I$. Hence, $\Bs(\Hs)$ (viewed
as acting on $\Hs$) is a factor (since $\Bs(\Hs)' = \CC I$), and the
only abelian factor is $\CC I$. Note that a von Neumann algebra $\Ms$
is abelian if and only if $\Ms \subset \Ms'$. $\Ms$ is said to be
{\it maximally abelian} (in $\Bs(\Hs)$) if $\Ms = \Ms'$. In this case,
the only abelian von Neuman algebra (in $\Bs(\Hs)$) containing $\Ms$
is $\Ms$ itself. Note that $L^\infty(X,\Ss,\mu)$ is maximally abelian
when acting on $L^2(X,\Ss,\mu)$.

     An immediate corollary of the double commutant theorem is that
the set of projections $\PM$ in a von Neumann algebra $\Ms$ is a
complete (orthomodular) lattice and that it determines $\Ms$
completely in the sense $\cM= \PM''$. This fact suggests that by
investigating the lattice $\PM$ one acquires insight into the
structure of the algebra itself. Indeed, Murray and von Neumann used
this lattice structure to begin a classification of von Neumann
algebras.  The key concept in the classification is the equivalence of
projections: two projections $A$ and $B$ in $\cM$ are called
{\it equivalent} ($A\sim B$) with respect to the algebra $\cM$ if there is
an operator (partial isometry) $W \in \cM$ which maps the range of
$I - A$ onto $\{ 0 \}$ and is an isometry between the ranges of
$A$ and $B$, \ie $W^* W = A$ and $WW^* = B$. The relation $\sim$
is an equivalence relation on $\PM$.
Let $\PM_{\sim}$ be the resultant set of equivalence classes. With the
help of $\sim$ one can introduce a partial ordering $\preceq$ on
$\PM$: $A\preceq B$ if there exists a $B' \in \Ps(\Ms)$ whose range is
contained in that of $B$ and which is equivalent to $A$, \ie $A\sim
B'\leq B$. A projection $A$ is called {\em finite} if it does not
contain any projection which is equivalent to $A$, \ie if $B\leq A$
and $B\sim A$ imply $A=B$. A projection is {\em infinite\/} if it is
not finite.

     From the point of view of Murray and von Neumann's classification
of von Neumann algebras, the important fact concerning $\preceq$ is the
{\it Comparison Theorem}: for any two $A,B\in\PM$ there exists a
projection $Z\in\PM\cap\PM'$ such that
$$ZAZ\preceq ZBZ  \mbox{\ \ and\ \ }(I- Z)B(I-Z)\preceq
(I-Z)A(I-Z) \, .$$
It follows that if $\cM$ is a factor, then $\PM_{\sim}$ is totally
ordered with respect to $\preceq$, \ie either $A\preceq B$, or
$B\preceq A$ holds for any $A,B \in \Ps(\Ms)$. Two factors cannot be
isomorphic as von Neumann algebras if the corresponding $\PM_{\sim}$
are not isomorphic as ordered spaces.

\begin{prop} \cite{MvN} \label{dimensionexistence}
If $\cM$ is a factor, then there exists a map
$d : \PM \rightarrow \lbrack 0, \infty\rbrack$, which is unique
up to multiplication by a constant and has the following properties:
\begin{description}
\item[(i)] $d(A)=0$ if and only if $A=0$.
\item[(ii)] If $A\c B$, then $d(A+B)=d(A)+d(B)$.
\item[(iii)] $d(A)\leq d(B)$ if and only if $A\preceq B$.
\item[(iv)] $d(A)<\infty$ if and only if $A$ is a finite projection.
\item[(v)]  $d(A)=d(B)$ if and only if $A\sim B$.
\item[(vi)] $d(A)+d(B)=d(A\es B)+d(A\vagy B)$.
\end{description}
\end{prop}
The map $d$ is called the {\em dimension function\/} on $\cP(\cM)$.
This proposition implies that the order type of $\PM_{\sim}$ can be
read off of the order type of the range of the function $d$. Murray
and von Neumann determined the possible ranges of $d$ in \cite{MvN}.
The result is shown in the table below (by choosing suitable
normalization of the function $d$).

\vskip0.5cm
\begin{tabular}{lll}
range of $d$ & type of factor $\cM$ &  \\
\hline
\hline
$\{0,1,2,\ldots n\}$ & I$_n$  & discrete, finite\\
\hline
$\{0,1,2,\ldots \infty\}$ & I$_{\infty}$  & discrete, infinite  \\
\hline
$\lbrack 0,1 \rbrack$ & II$_{1}$  & continuous, finite \\
\hline
$\{x\quad \vert 0\leq x \leq \infty \}$  & II$_{\infty}$&continuous,
infinite \\
\hline
$\{0,\infty \}$ & III &  purely infinite                     \\
\hline
\end{tabular}
\vskip0.5cm

     This thus results in a classification of von Neumann algebras
into type I$_n$, I$_{\infty}$, II$_{1}$, II$_{\infty}$ and III. Any
von Neumann algebra can be decomposed into a direct sum of algebras of
these types. And every factor is exactly one of these types.  As
tensor products play an important role in quantum theory, it is useful
to know that if $\Ms$ is type I$_n$ and $\Ns$ is type I$_m$, then
their tensor product\footnote{the von Neumann algebra on $\Hs \otimes \Hs$
generated by operators of the form $M \otimes N$, $M \in \Ms$ and $N \in 
\Ns$.}
$\Ms \otimes \Ns$ is type I$_{nm}$.  If $\Ms$ and
$\Ns$ have no direct summand of type III and one of them is type II
(\ie has only direct summands of type II$_1$ or type II$_\infty$),
then $\Ms \otimes \Ns$ is type II. And if $\Ms$ is type III, then
so is $\Ms \otimes \Ns$, for any von Neumann algebra $\Ns$. Note
further that $\Ms$ is type I (resp. II, III) if and only if
$\Ms'$ is type I (resp. II, III).

     The algebra $\Bs(\Hs)$ is of type I$_n$ if the dimension of $\Hs$
is $n$ and is of type I$_\infty$ if $\Hs$ is infinite dimensional.
Moreover, a von Neumann algebra $\Ms$ is of type I if and only if it
is isomorphic to $\Bs(\Hs) \otimes \As$, for some Hilbert space $\Hs$,
where $\As$ is some abelian von Neumann algebra. Hence, all abelian
von Neumann algebras are of type I. But there are other types of von
Neumann algebras, and these other types have properties {\it
radically} different from the properties of $\BH$. For instance:

\begin{enumerate}
\item If $\cM$ is finite, {\it i.e.} it has only direct summands of type
I$_n$ or II$_1$,  then its projection lattice $\PM$ is
modular; whereas if $\Ms$ is infinite, $\PM$ is orthomodular but not
modular.

\item $\cM$ is of type I (\ie it has only direct summands of type
I$_n$ or I$_{\infty}$) if and only if it has nonzero minimal
projections. (These are atoms in the projection lattice $\Ps(\Ms)$.)

\item There exists a faithful normal tracial state on a factor $\Ms$ if and
only if $\Ms$ is finite.

\item If $\cM$ is of type III, then all of its nonzero projections are
infinite. This implies that for any projection $P$ in a type III
algebra there exist countably infinitely many
mutually orthogonal projections $P_i \in \Ms$ such that $P=\vagy P_i$.
\end{enumerate}

     After the breakthrough of the Tomita--Takesaki modular theory at
the end of the 1960's \cite{Takm}, it became possible for A. Connes
\cite{Con} to further refine the classification of type III algebras
into an uncountably infinite family of type III$_\lambda$ algebras,
$\lambda \in [0,1]$. Particularly the type III$_1$
case is of physical interest, as will be seen in Sections
\ref{necessity} and \ref{differences}.

     The distinction between types of von Neumann algebras will be
seen to have consequences for probability theory, but first we must
explain where the probability theory is to be found in this
structure. To this end, it will be useful to review Kolmogorov's
formulation of classical probability theory from the vantage point of
operator algebra theory.

\section{Classical Probability Theory} \label{classical}

     A classical {\it probability space} is a triplet $(X,\cS,p)$, where $X$
is a set, $\cS$ is a $\sigma$-algebra of subsets of $X$, and
$p\colon\cS\to\lbrack 0,1\rbrack$ is a $\sigma$-additive measure,
\ie for every countable collection
$\{ S_n \}_{n \in \IN} \subset \Ss$ of mutually disjoint measurable
sets, one has
\begin{equation} \label{additivity}
p(\bigcup_{n \in \IN} S_n) = \sum_{n \in \IN} p(S_n) \, .
\end{equation}
The elements $S \in \Ss$ are interpreted as possible events and $p(S)$
as the probability that event $S$ takes place. The probabilities for a
suitable subclass of events $S$ are the primary data, and the measure
$p$ on a suitably generated $\sigma$-algebra $\Ss$ is generally a
derived quantity.\footnote{This may be seen in standard books on
measure theory such as \cite{Hal}.}

     Another crucial concept needed in physical applications of
probability theory is the concept of random variable, which
is used to represent the observable physical quantities in concrete
applications. A map $f\colon X\to \bbbr$ is a (real-valued
Borel measurable) {\it random variable} if $f^{-1}(B)\in\cS$,
for all $B\in\cB(\RR)$, where $\cB(\bbbr)$ is the Borel $\sigma$-algebra
of $\bbbr$. A distinguished subclass of such random variables is
constituted by the essentially bounded measurable functions
$f \in L_\RR^\infty(X,\cS,p)$. It is often convenient to view
$L_\RR^\infty(X,\cS,p)$ as a subset of the space of complex-valued
essentially bounded measurable functions $L^\infty(X,\cS,p)$. Within
this subset is the subclass
$\Ps(\Ss) = \{ \chi_S : S \in \Ss \}$ of characteristic functions:
$$\chi_S(x) = \left\{ \begin{array}{ll} 1 & \mbox{if $x \in S$} \\
                                    0 & \mbox{otherwise}
\end{array} \right. \, ,$$
for which $p(S) = \int_X \chi_S(x) \, dp(x) $. Viewed as multiplication
operators on $L^2(X,\Ss,p)$, each characteristic function $\chi_S$ is
an orthogonal projection, and the linear span of $\Ps(\Ss)$ is
dense in the von Neumann algebra $L^\infty(X,\cS,p)$, \ie $\Ps(\Ss)$
generates $L^\infty(X,\cS,p)$. Indeed, $\Ps(\Ss)$ coincides with the
set of all projections in the von Neumann algebra $L^\infty(X,\cS,p)$.

     Therefore, classical probability theory yields a distinguished
Hilbert space $L^2(X,\Ss,p)$ on which acts a distinguished abelian
von Neumann algebra $L^\infty(X,\cS,p)$ generated by the set $\Ps(\Ss)$
of its projections, each of which has significance in the given
probabilistic model. The probability measure $p$ determines uniquely
a state $\phi$ on $L^\infty(X,\cS,p)$ by
$$\phi(f) = \int_X f(x) \, dp(x) \, ,$$
for all $f \in L^\infty(X,\cS,p)$. And because the measure is
$\sigma$-additive, this state is normal on $L^\infty(X,\cS,p)$ --- cf.
(\ref{normal}) and (\ref{additivity}), since $p(S) = \phi(\chi_S)$,
for all $S \in \Ss$, and $\chi_{S_1} \cdot \chi_{S_2} = 0$ if and only if
$S_1 \cap S_2 = \emptyset$, modulo sets of $p$-measure zero. The
probabilistically fundamental data, $p(S)$ with $S \in \Ss$, are
reproduced by the expectations, $\phi(P)$ with
$P \in \Ps(L^\infty(X,\cS,p)) = \Ps(\Ss)$, of the projections from the
von Neumann algebra in the state $\phi$.

     Of course, there are further, derived structures in classical
probability theory, and some of these are discussed in Section
\ref{differences}.

\section{Noncommutative Probability Theory} \label{quantum}

     We have seen in the preceding section that classical probability
theory yields an abelian von Neumann algebra with a specified normal
state.  On the other hand, we have also seen in Section \ref{algebra}
that an abelian von Neumann algebra with a specified normal state
yields a measure space with a probability measure on it, which is
precisely the point of departure in Kolmogorov's formulation of
probability theory. Although these observations are remarkable, why do
operator algebraists repeat the phrase ``von Neumann algebra theory is
noncommutative probability theory''\footnote{or noncommutative measure
theory} like a mantra? We cannot hope to explain here the full scope
of noncommutative probability theory\footnote{which already subsumes
noncommutative generalizations of many basic and advanced concepts of
classical probability theory, from conditional expectations and
central limit theorems to the theory of stochastic processes and
stochastic integration, as well as notions from measure theory such as
Radon-Nikodym derivatives, Lebesgue decomposition and $L^p$ spaces ---
cf. \cite{Par,Mey,Str,CO,Hol,Ham} for further reading}.
Instead, we must content ourselves with highlighting some of those
aspects which are of particular importance to quantum theory.

     From the beginning of quantum theory, orthogonal projections have
played a central role, whether it be logical, operational or
mathematical. They are used in the description of what are called
``yes--no experiments'' --- is the spin of the electron pointed in
this spatial direction? --- is the particle to be found in this subset
of space? --- is the atom in its ground state? Moreover, through von
Neumann's spectral theorem, general observables could be constructed
out of these particularly elementary observables: even for an
unbounded self-adjoint operator $A$ on $\Hs$, there exists a measure
$\nu$ on the spectrum $\sigma(A)$ of $A$ taking values in
$\Ps(\Bs(\Hs))$ such that
\begin{equation} \label{spectraltheorem}
A = \int_{\sigma(A)} \lambda \, d\nu(\lambda) \, ,
\end{equation}
where the convergence is in the strong operator topology on the domain
of $A$. Later, it was proven that if $\Ms$ is any von Neumann algebra
and $A = A^* \in \Ms$, then there exists a measure $\nu$ on the
spectrum $\sigma(A)$ of $A$ taking values in $\PM$ such that equation
(\ref{spectraltheorem}) holds, thus establishing another sense in
which $\Ms$ is generated by $\PM$.

     Indeed, one has a Borel functional calculus in arbitrary von
Neumann algebras.

\begin{prop} \cite[Theorem 5.2.8]{KR}  \label{functionalcalculus}
Let $\Ms$ be a von Neumann algebra and $\Fs$ be the *-algebra of bounded
complex-valued Borel measurable functions on $\CC$. Let $A \in \Ms$ be
self-adjoint\footnote{It actually suffices that $A$ be normal.} and
$f \in \Fs$. Then
$$f(A) \equiv \int_{\sigma(A)}f( \lambda) \, d\nu(\lambda) \, ,$$
is an element of $\Ms$, and the map $f \mapsto f(A)$ is a normal
*-homomorphism from $\Fs$ into $\Ms$. The image of $\Fs$ under this
map is the abelian subalgebra $\{ A \}''$ of $\Ms$.
If $f$ vanishes identically
on $\sigma(A)$, then $f(A) = 0$. One has $\overline{f}(A) = f(A)^*$
and $(f \circ g)(A) = f(g(A))$, for any $f,g \in \Fs$. Moreover,
the mapping $S \mapsto \chi_S(A) \in \PM$ yields a projection-valued
$\sigma$-additive measure on the Borel subsets of $\CC$ such that
$A$'s spectral resolution $\{ E_\lambda \}_{\lambda \in \RR}$ is
given by $E_\lambda = I - \chi_{(\lambda,\infty)}(A)$.\footnote{Recall
that $\nu$ can be recovered from $\{ E_\lambda \}$.}
\end{prop}

\noindent Computations one is already familiar with in $\Bs(\Hs)$ can
therefore be performed in arbitrary von Neumann algebras. And
self-adjoint elements $A \in \Ms$ yield natural $\sigma$-algebra
homomorphisms $A : \Bs(\RR) \rightarrow \PM$ in analogy to random
variables.

     The projections in $\Ms$ have also been interpreted as yes--no
propositions in a propositional calculus with a view towards
establishing a quantum logic or finding another foundation for quantum
theory. This idea first appeared in the seminal paper by G.D. Birkhoff
and von Neumann \cite{BvN}, where it was assumed that the type II$_1$
algebras play a privileged role in quantum theory (see \cite{Red1} for
an analysis of this early concept of quantum logic).  Subsequent
developments have shown, however, that the type II$_1$ assumption is
too restrictive --- cf. \cite{Mac,Jau,Red}. 
From this point of view, a normal state $\phi$ on $\Ms$ provides
an interpretation (in the sense of logic) of the quantum propositional 
calculus, an interpretation which satisfies (\ref{normal}).

     For whichever reason one accepts the basic nature of the
projections in a von Neumann algebra $\Ms$, any normal state
$\phi$ on $\Ms$ provides a map
$$\PM \ni P \mapsto \phi(P) \in [0,1] \, , $$
which is $\sigma$-additive in the sense of (\ref{normal}). From the
discussion in Section \ref{classical}, it can now be seen that
(\ref{normal}) is the noncommutative generalization of
(\ref{additivity}). Thus, every normal state on $\Ms$ determines a
$\sigma$-additive probability measure\footnote{note: $\phi(I) = 1$} on
the lattice $\PM$. For mathematical, operational and logical reasons,
the converse of this relation became pressing: given a map
$\mu : \PM \rightarrow [0,1]$ satisfying (\ref{normal}),\footnote{so,
necessarily, $\mu(I) = 1$} does there exist a normal state $\phi$
on $\Ms$ such that $\phi(P) = \mu(P)$ for all $P \in \PM$? Gleason
\cite{Gle} showed that if $\Ms = \Bs(\Hs)$ and the dimension of $\Hs$
is strictly greater than 2, then this converse is indeed true. And
in a lengthy effort, to which there were many contributors (see
\cite{Mae,Ham} for an overview of this development, as well as proofs),
it was shown that Gleason's Theorem could be generalized to (nearly)
arbitrary von Neumann algebras.

\begin{prop}  \label{Gleason}
Let $\Ms$ be a von Neumann algebra with no direct summand of type
I$_2$. Then every map $\mu : \PM \rightarrow [0,1]$ satisfying
(\ref{normal}) extends uniquely to a normal state on $\Ms$.
Moreover, every finitely additive\footnote{\ie (\ref{normal}) holds for
finite families of mutually orthogonal projections} map
$\mu : \PM \rightarrow [0,1]$ extends uniquely to a state on $\Ms$.
\end{prop}

\noindent This theorem makes clear that any probability theory based upon
suitable lattices of projections in Hilbert spaces is subsumed in
the framework of normal states on von Neumann algebras.

     In noncommutative probability theory, a probability space is a
triple $(\Ms,\PM,\phi)$ consisting of a von Neumann algebra, its
lattice of orthogonal projections and a normal state on the algebra.
As is now clear, the classical starting point is regained precisely
when $\Ms$ is abelian. Examples of derived notions in probability
theory with generalization in the noncommutative theory --- independence
and conditional expectations --- are discussed in Section
\ref{differences}.

     There are many significant differences between (truly)
noncommutative probability theory and classical probability theory. A
state $\phi$ on a von Neumann algebra $\Ms$ is said to be {\it
dispersion free} if $\phi(A^2) - \phi(A)^2 = 0$, for all $A = A^* \in
\Ms$. Of course, if $\Ms$ is abelian, then it admits many dispersion
free states (pure states, which are multiplicative on abelian \C
algebras, \ie $\phi(AB) = \phi(A)\phi(B)$, for all $A,B \in \Ms$). But
a nonabelian factor admits no dispersion free states
\cite[Cor. 2]{Mis}. That $\Ss$ is a Boolean algebra and $\PM$ is not
when $\Ms$ is nonabelian is another crucial difference. The
consequences of these two differences alone have generated a
voluminous literature.  This is not the place to discuss these matters
in further detail.  However, some other important differences {\it
are} discussed in Section \ref{differences}.

\section{Quantum Mechanics: Type I Noncommutative Probability Theory}
\label{QM}

     In light of what has been presented above, one now sees that the
basic components of the noncommutative probability theory are inherent
in nonrelativistic quantum mechanics. The unit vectors in and the
density matrices on $\Hs$, which are used to model the preparation of
the quantum system, induce normal states on $\Bs(\Hs)$. The
observables of the system are modelled by self-adjoint elements of
$\Bs(\Hs)$. And the expectation of the observable $A$ of a system
prepared in the state $\phi$ is given by $\phi(A)$. Of course,
quantum theory supplements this basic framework with further
structures.

     In nonrelativistic quantum mechanics, the operator algebras which
arise in modelling algebras of observables are exclusively type I
algebras. Typically, they are either $\Bs(\Hs)$ for some Hilbert space
$\Hs$, or they are abelian subalgebras of $\Bs(\Hs)$ generated by a
commuting family of observables (see below) or perhaps by a single
observable (cf. Prop. \ref{functionalcalculus}). Characteristic of
nonrelativistic quantum mechanics is the restriction to systems
involving only finitely many degrees of freedom. The Hilbert space
$\Hs$ may be finite or infinite dimensional, depending on which
observables are being employed. For example, the use of position and
momentum observables entails that $\Hs$ be infinite dimensional (since
there is no representation of the canonical commutation relations in a
finite dimensional Hilbert space), while finite dimensional Hilbert
spaces suffice when considering only observables associated with spin.
In the former case, an irreducible representation of the canonical
commutation relations is normally used (cf. \cite{Sum01}), in which
case the von Neumann algebra generated by the spectral projections of
the position and momentum operators is $\Bs(\Hs)$. In the latter case,
the components of the spin generate the full matrix algebra
$\Bs(\Hs)$.

     But abelian algebras are also frequently employed in quantum
mechanics. Two observables $A,B$ are said to be
compatible (or commensurable) if they commute: $[A,B] = AB - BA = 0$.
Any family $\{ A_\alpha \}$ of mutually commuting (bounded)
self-adjoint operators acting on $\Hs$ generates an abelian von
Neumann algebra $\{ A_\alpha \}''$, which is thus isomorphic to a
suitable algebra of bounded Borel functions. It is then possible to
construct a self-adjoint $A \in \Bs(\Hs)$ and bounded Borel functions
$f_\alpha$ such that $A_\alpha = f_\alpha(A)$, for all $\alpha$.
Hence, when dealing only with compatible observables, classical
probability suffices. Note that the complete commuting families of
observables, which play an important role in some parts of quantum
theory, are precisely those families $\{ A_\alpha \}$ for which
$\{ A_\alpha \}''$ is maximally abelian (viewed as a subalgebra of
$\Bs(\Hs)$).

\section{The Necessity of Non-Type--I Probability Spaces in Physics}
\label{necessity}

     The different types of noncommutative probability spaces
described in Section \ref{algebra} are not mere mathematical
curiosities --- they are indispensable in physical applications,
because certain quantum systems {\em cannot\/} be described using only
type I algebras.  Such quantum systems typically have infinitely many
degrees of freedom and arise in quantum statistical mechanics and in
relativistic quantum field theory. We shall briefly indicate how this
occurs.

\subsection{Quantum Statistical Mechanics} \label{qstatphys}

     Although the physical systems described by quantum statistical
mechanics actually have only finitely many degrees of freedom, the
number of degrees of freedom is so large that physicists have found it
to be more convenient to work with the idealization known as the
thermodynamic limit (or infinite volume limit) of such systems.  In
this limit, the number of degrees of freedom is indeed infinite.  The
simplest nontrivial examples of such systems include the so-called
lattice gases, which are mathematically precise models of discrete
quantum statistical mechanical systems in the thermodynamic limit.

\subsubsection{The quasilocal structure of lattice gases}  \label{structure}

     The simplest example of a lattice gas is the one dimensional
lattice gas. A quantum system (typically representing an atom or
molecule) is located at each point $i$ on a one dimensional lattice
$\cZ$ (taken to be the additive group of integers) infinite in both
directions:
$$
\ldots\qquad\stackrel{\cB(\cH_{i-2})}{\stackrel{\bullet}{ i-2
}}\quad \stackrel{\cB(\cH_{i-1})}{\stackrel{\bullet}{\textstyle i-1
}}\quad \stackrel{\cB(\cH_{i})}{\stackrel{\bullet}{\textstyle i}}\quad
\stackrel{\cB(\cH_{i+1})}{\stackrel{\bullet}{\textstyle i+1}}\quad
\stackrel{\cB(\cH_{i+2})}{\stackrel{\bullet}{\textstyle i+2
}} \qquad \ldots
$$
where the $\cH_i$ are copies of a fixed finite dimensional Hilbert
space $\Ks$. The self-adjoint elements of $\cB(\cH_i)$ represent the
observables of the system at site $i$. For a {\em finite\/} subset
$\Lambda\subset\cZ$, the tensor product\footnote{As the \C algebras
$\Bs(\Hs_i)$ are von Neumann algebras, the tensor product here may
be taken to be the (unique) spatial tensor product, as is done throughout
this paper.}
\begin{equation}
\cA(\Lambda)=\otimes_{i\in\Lambda}\cB(\cH_i)
\end{equation}
represents a quantum system localized in region $\Lambda$. If
$\Lambda_1 \subset \Lambda_2$, the algebra
$\cA(\Lambda_1)$ is isomorphic to and is identified with the algebra
$\otimes_{i \in \Lambda_2} \, \Bs_i \subset \cA(\Lambda_2)$, where
$\Bs_i = \CC I_i$, for $i \notin \Lambda_1$, and $\Bs_i = \Bs(\Hs_i)$,
for $i \in \Lambda_1$. Then, the collection
$\{ \cA(\Lambda) : \Lambda \subset \Zs \; \textnormal{finite} \}$
is a directed set under inclusion, and the union
$\cA_0= \cup_{\Lambda}\cA(\Lambda)$ is an incomplete normed algebra
with involution, which contains all observables which can be localized in
some finite region. The minimal norm completion $\cA$ of $\cA_0$ is
then a \C algebra \cite{Sak}. The self-adjoint elements of this \C algebra
represent the observable quantities of the infinitely extended quantum
system. 

     The one dimensional lattice gas also possesses a natural spatial
symmetry --- translations along the lattice. Since all Hilbert spaces
$\cH_i$ are identical copies of the same Hilbert space $\cK$, there is
a unitary operator $U_i(j)$ which takes the Hilbert space $\cH_i$ from
site $i$ to site $i+j$:
\begin{equation}
U_i(j)\colon \cH_i\to\cH_{i+j}
\end{equation}
These unitaries may be chosen to possess the group property
\begin{equation}
U_i(j+l)=U_{i+l}(j)U_i(l) \, ,
\end{equation}
and they transform observables localized at individual sites:
\begin{equation}
\cB(\cH_i)\ni A\mapsto U_i(j)AU_i(j)^{-1} \in \cB(\cH_{i+j}) \, .
\end{equation}
Taking the products $U_{\Lambda}(j)=\otimes_{i\in\Lambda}U_i(j)$, one
obtains unitaries $U_{\Lambda}(j)$ which take $\cH_{\Lambda}$ onto
$\cH_{\Lambda+j}$\footnote{$\Lambda+j = \{ i+j : i \in \Lambda \}$}
and which act upon the local observables:
\begin{equation}
\cB(\cH_{\Lambda})\ni A\mapsto U_{\Lambda}(j)AU_{\Lambda}(j)^{-1} \in
\cB(\cH_{\Lambda+j}) \, .
\end{equation}
Defining
$$\alpha_j(A) = U_{\Lambda}(j)AU_{\Lambda}(j)^{-1} \, , \,
A \in \As(\Lambda) \, , \, {\rm finite} \, \Lambda \subset \Zs \, , $$
yields an automorphism on $\As_0$ which extends to an
automorphism $\alpha_j$ of $\cA$. Thus one obtains a representation
$\cZ\ni i \mapsto \alpha_i \in \rm{Aut}(\As)$ of the translation symmetry
group of the infinite lattice gas by automorphisms on $\cA$.

     Also a time evolution of the infinite lattice gas can be
constructed from the time evolutions of its local systems. If
$H_{\Lambda}$ is the generator of the unitary group $U_t$,
$t\in\bbbr$, which gives the time evolution of the quantum system
localized in $\Lambda$ (so that $H_{\Lambda}$ carries the
interpretation of the total energy operator for the subsystem in
$\Lambda$), then (under suitable assumptions \cite{BRII}) the adjoint
action of $U_t$ can be extended in a similar manner to an automorphism
$\alpha_t$ of $\cA$. The dynamical behavior of the infinite lattice
gas is then encoded in the {\em \C dynamical system\/}
$(\cA,\{\alpha_t\}_{t\in\bbbr})$, where $t\mapsto\alpha_t$ is a
continuous representation of $(\bbbr,+)$ by automorphisms of
$\cA$.\footnote{Note that only if the local Hamiltonian operators,
$H_{\Lambda}$, are bounded will the resultant representation
be continuous in the sense required by the term \C dynamical system:
the map $\RR \ni t \mapsto \alpha_t(A) \in \As$ must be continuous for
all $A \in \As$. For unbounded energy operators, the representation
is continuous in a weaker sense. The results stated
below can be extended in a suitable manner to the latter case, as
well, but these technicalities will be suppressed here.}

     This mathematical model of the one dimensional lattice gas can
thus be compactly summarized as follows: There exists a net of local
algebras of observables
$$\cZ\supset \Lambda \mapsto \As(\Lambda) = \cB(\cH_{\Lambda}) \, ,$$
indexed by the finite subsets $\Lambda$ of the lattice $\cZ$, with
these properties:
\begin{enumerate}
\item {\bf Isotony}: If $\Lambda_1 \subseteq \Lambda_2$, then
$\cA(\Lambda_1)$ is a subalgebra of $\cA(\Lambda_2)$. This then
enables one to define the {\it quasilocal algebra}:
$$\cA = \overline{\cup_{\Lambda \subset \cZ, \rm{finite}} \,
\cA(\Lambda)}^{norm} \, .$$

\item {\bf Local commutativity}: If
$\Lambda_1 \cap \Lambda_2 = \emptyset$, then $[A_1,A_2]=0$, for all
$A_1 \in \cA(\Lambda_1)$ and $A_2 \in \cA(\Lambda_2)$.

\item {\bf Covariance}: There is a representation of the
symmetry group $(\cZ,+)$ of the space $\cZ$ by automorphisms
$\alpha$ on $\cA$ such that
$$\alpha_a\cA(\Lambda) = \cA(\Lambda+a) \, ,$$
for all $a \in \Zs$ and finite $\Lambda \subset \Zs$.

\item {\bf Time evolution}: The dynamical behavior of the
system is given by a continuous representation of $(\bbbr,+)$
by automorphisms of $\cA$, yielding a \C dynamical system
$(\As,\{\alpha_t\}_{t\in\bbbr})$.
\end{enumerate}

\subsubsection{States on the Quasilocal Algebra of the Lattice Gas}
\label{states}

     A fundamental task of investigating quantum statistical systems
is the determination of their equilibrium states. For a limited
class of quantum systems modelled by a \C dynamical system with
a total energy operator $H$, the
usual Gibbs equilibrium state at inverse temperature $\beta$,
$$\phi(A) = \frac{\tr(e^{-\beta H}A)}{\tr(e^{-\beta H})} \, , \,
A \in \As \, ,$$
is entirely satisfactory. However, in most applications to systems
involving infinitely many degrees of freedom, the operator
$e^{-\beta H}$ is not of trace class, and the Gibbs state is not defined.
Through the efforts of a number of leading mathematical physicists, it has
been well established that the appropriate notion of equilibrium state
of a \C dynamical system is that of a KMS state --- cf. \cite{BRII}:

\begin{definition}
A state $\phi$ on the \C algebra $\cA$ of the \C dynamical system
$(\cA,\{\alpha_t\}_{t\in\bbbr})$ is an $(\alpha,\beta)$-KMS state at
inverse temperature $\beta\in\bbbr$, if for every $A,B \in \As$ there
exists a complex-valued function $f_{A,B}$ which is analytic in the strip
$\{ z \in \CC : 0 < {\rm Im}(z) < \beta \}$ and continuous on the
closure of this strip such that
$$f_{A,B}(t + i0) = \phi(\alpha_t(A)B)  $$
and
$$f_{A,B}(t + i\beta) = \phi(B\alpha_t(A)) \, ,$$
for all $t \in \RR$.
\end{definition}

     Note that Gibbs states are KMS states. Moreover, any KMS state
$\phi$ on a \C dynamical system $(\As,\{\alpha_t\}_{t\in\bbbr})$ is
{\it $\alpha$-invariant}, \ie $\phi \circ \alpha_t = \phi$, for all
$t \in \RR$. Another characteristic feature
of KMS states, which is of particular relevance to our considerations
here, is spelled out in the next proposition.

\begin{prop}\cite{PW}\cite[Corollary 5.3.36]{BRII} \label{KMSIII}
If $\phi$ is a KMS state of the \C dynamical system
$(\cA,\{\alpha_t\}_{t\in\bbbr})$ with $\beta > 0$
(and \eg $\phi$ is weakly clustering),
then $\pi_{\phi}(\cA)''$ is a type III factor.
\end{prop}
Hence, one cannot describe phenomena such as equilibrium (thus also phase
transition) in quantum physics without going beyond the type I von
Neumann algebras.

     A special case of $(\alpha,\beta)$-KMS states is the infinite
temperature ($\beta=0$) KMS state. Such states are called chaotic,
and an $(\alpha,0)$-KMS state is an $\alpha$-invariant tracial state
(and {\it vice versa}). In
the case of the one dimensional lattice gas, tracial states can be
constructed explicitly. If $\widehat{\tr}_i$ is the normalized trace on
$\cB(\cH_i)$ and $\tau_{\Lambda}$ is defined on
$\otimes_{i\in\Lambda}\cB(\cH_i)$ by
\begin{equation}
\tau_{\Lambda}(A_{i_1}\otimes A_{i_2}\ldots \otimes A_{i_k}) =
\widehat{\tr}_{i_1}(A_{i_1}) \, \widehat{\tr}_{i_2}(A_{i_2})
\cdots \widehat{\tr}_{i_k}(A_{i_k})
\, ,
\end{equation}
with $\Lambda=\{i_1,i_2,\ldots, i_k\}$, then $\tau_0$ defined by
$\tau_0(A)=\tau_{\Lambda}(A)$, for $A\in\cA(\Lambda)$ and finite
$\Lambda \subset \Zs$, yields a
norm-densely defined linear functional on $\cA$. The extension
$\tau$ of $\tau_0$ from $\cA_0$ to $\cA$ is a tracial state.
As $\cA$ is a simple algebra, it may be identified with its GNS
representation associated with the state $\tau$. Let $\Hs$
denote the corresponding Hilbert space. 

     It follows that the algebra $\cA$ representing the bounded
observables of the infinitely extended lattice gas cannot
be $\BH$, because there is no tracial state on $\BH$ ($\cH$ is
infinite dimensional, since $\Zs$ is infinite).  The state $\tau$
induces a tracial state on the von Neumann algebra $\cA''$;
so $\cA''$ cannot be type I, either. In fact, the von
Neumann algebra $\cA''$ is a type II$_1$ von Neumann algebra
(see Section XIV.1 in \cite{Tak}). This is typical of GNS representations
associated with chaotic states.

     Another type of physically relevant state is the ground state,
which formally is an $(\alpha,\infty)$-KMS state. If a state $\phi$ on
a \C dynamical system $(\As,\{\alpha_t\}_{t\in\bbbr})$ is
$\alpha$-invariant, then in the corresponding GNS space there exists a
self-adoint operator $H_\phi$ such that
\begin{equation}  \label{ground}
e^{itH_\phi} \pi_\phi(A) e^{-itH_\phi} = \pi_\phi(\alpha_t(A)) \, ,
\, A \in \As \, , \, t \in \RR \, .
\end{equation}
A state $\phi$ on a \C dynamical system $(\As,\{\alpha_t\}_{t\in\bbbr})$
is a {\it ground state}, if $\phi$ is $\alpha$-invariant and $H_\phi \geq
0$.
For ground states one has the following result.

\begin{prop} \cite[Theorem 5.3.37]{BRII}
If $\phi$ is an (extremal) ground state of a \C dynamical
system $(\As,\{\alpha_t\}_{t\in\bbbr})$, then $\pi_{\phi}(\cA)''$ is
a type I factor.
\end{prop}

     In sum, the behavior of general quantum systems modelled by \C
dynamical systems $(\cA,\alpha)$ cannot be described using solely type
I algebras, which typically only arise (for the quasilocal algebra $\As$)
in the GNS representation corresponding to a ground state. Note, however,
that in the simple lattice gas models discussed above, the local
algebras of observables, $\As(\Lambda)$, {\it are} type I algebras.
This is not typical of quantum statistical models, as shall be seen below.

\subsection{Brief Return to General Quantum Statistical Mechanics}
\label{brief}

     If one wishes to model quantum gases on suitable lattices in two
or three dimensional space, then one can repeat the steps described
above and arrive
at the same sort of quasilocal structure. And more general
models on such lattices can be constructed which still manifest the
structure properties 1-4 emphasized at the end of Section
\ref{structure}.  One also can drop the assumption that the physical
systems are restricted to lattices and consider an assignment of
suitable algebras $\cA(\Lambda)$ of observables to bounded regions
$\Lambda$ in the three dimensional Euclidean space $\bbbr^3$ -- the
resulting structure describes the thermodynamical limit of quantum
systems in three dimensional space. The Euclidean group replaces the
lattice symmetry group in such cases. This also leads to structures
entirely analogous to those manifested by the lattice gas models ---
cf. \cite{Emc,BRII,Sew}. The theorems discussed in Section
\ref{states} are therefore applicable to such models, as well. Note
that there are models of this kind in which even the local observable
algebras $\As(\Lambda)$ are of type III or type II 
(see \eg \cite{AWo,AWy}).

     {\it Local quantum physics} is the name given to the branch of
mathematical physics which investigates the mathematical models of
quantum systems wherein taking account of the localization of observables
leads to structures with properties analogous to those isolated at
the end of Section \ref{structure}. This approach has also
proven to be fruitful in the study of relativistic quantum fields on
general space--times.

\subsection{Local Relativistic Quantum Field Theory} \label{QFT}

     Here we restrict our attention to quantum fields on four
dimensional Minkowski space $M$, so the label $\Lambda$ (replaced by
convention with $\Os$ in quantum field theory) indicates the
spatio{\em temporal} localization of the algebra $\cA(\Os)$ of
observables in $M$. The algebra $\As(\Os)$ is interpreted as the
algebra generated by all the observables measurable in the spacetime
region $\Os$. The net of local algebras of
observables
\begin{equation}
\Os \mapsto \cA({\Os})
\end{equation}
indexed by the open bounded spacetime regions $\Os$ of the Minkowski
space--time is assumed to satisfy a number of physically motivated
conditions (cf. \cite{Haa,Ara}), which closely resemble the structural
properties of the net describing the one dimensional lattice gas and
which are natural in light of the mentioned interpretation.

\begin{enumerate}
\item {\bf Isotony}: If $\Os_1\subseteq \Os_2$, then $\cA(\Os_1)$
        is a subalgebra of $\cA(\Os_2)$. This enables the definition
of the quasilocal algebra as the inductive limit of the net, \ie the
smallest \C algebra $\cA$ containing all the local algebras $\cA(\Os)$.

\item {\bf Local commutativity (Einstein causality)}: If $\Os_1$ is
spacelike
separated from $\Os_2$, then $[A_1,A_2]=0$, for all $A_1 \in \cA(\Os_1)$
and $A_2 \in \cA(\Os_2)$.

\item {\bf Relativistic covariance}: There exists a continuous
representation $\alpha$ of the identity-connected component $\Pid$
of the Poincar\'e group by automorphisms on $\cA$ such that
$\alpha_\lambda(\cA(\Os))=\cA(\lambda\Os)$, for all
$\Os$ and $\lambda\in\Pid$.

     Though there are many kinds of physically relevant
representations, one of the most completely studied is the {\it vacuum
representation}.

\item {\bf Irreducible vacuum representation}: For each $\Os$,
$\cA(\Os)$ is a von Neumann algebra acting on a separable Hilbert space
$\cH$ in which $\cA'' = \cB(\cH)$, in which there
is a distinguished unit vector $\Omega$, and on which
there is a strongly continuous unitary representation $U(\Pid)$ satisfying
$U(\lambda)\Omega = \Omega$ , for all $\lambda \in \Pid$, and
$$\alpha_\lambda(A) = U(\lambda)AU(\lambda)^{-1} \quad , \quad
{\rm for \; all} \quad A \in \cA \quad ,$$
as well as the spectrum condition: the spectrum of the self-adjoint
generators of the strongly continuous unitary representation $U(\bbbr^4)$
of the translation subgroup of $\Pid$ (which has the physical
interpretation of the global energy--momentum spectrum of the theory)
must lie in the closed forward light cone.

    A common assumption made when dealing with vacuum representations
is given next.

\item {\bf Weak additivity}: For each $\Os$,
$$\{ U(x) \As(\Os) U^{-1}(x) : x \in \RR^4 \}'' = \As''
\, .$$
\end{enumerate}

     These assumptions entail the Reeh--Schlieder Theorem
(cf. \cite{Haa,Ara}), which permits the use of Tomita--Takesaki
modular theory in quantum field theory \cite{Haa,Ara}.

\begin{prop} For every $\Os$, the vector $\Omega$ is cyclic and separating
for $\As(\Os)$, \ie the set of vectors $\As(\Os)\Omega$ is dense in $\Hs$,
and $A \in \As(\Os)$ and $A\Omega = 0$ entail $A = 0$.
\end{prop}

\noindent Thus, no local observable can annihilate $\Omega$, \ie
$\langle \Omega, P \Omega \rangle \neq 0$, for all projections
$P \in \As(\Os)$.

     Many concrete models satisfying these conditions have been
constructed, though none of them is an interacting quantum field in
four spacetime dimensions. Of course, no such model has {\it ever}
been constructed,\footnote{In two and three spacetime dimensions,
models of interacting quantum fields satisfying conditions 1--5 have
been constructed --- see \eg \cite{GJ,Sch,Sum82}.} so one can hardly
attribute the source of the problem to the set of ``axioms'' above. On
the contrary, we are convinced that the above conditions are
operationally natural and express the minimal conditions to be
satisfied by any local relativistic quantum field theory in the vacuum
on Minkowski space. So we view consequences of these assumptions to be
generic properties in the stated context.

     It will be convenient to concentrate attention on two special
classes of spacetime regions in $M$. A double cone is a (nonempty)
intersection of some open forward light cone with an open backward
light cone. Such regions are bounded, and the set $\Ds$ of all double
cones is left invariant by the natural action of $\Pid$ upon it. An
important class of unbounded regions is specified as follows.
After choosing a coordinatization of $M$, one defines the right wedge
to be the set
$W_R = \{ x = (x_0,x_1,x_2,x_3) \in M : x_1 > \vert x_0 \vert \}$
and the set of wedges to be $\Ws = \{ \lambda W_R : \lambda \in \Pid \}$.
The set of wedges is thus independent of the choice of coordinatization;
only which wedge is called the right wedge is coordinate-dependent.
Because $\Ds$ is a base and $\Ws$ is (nearly \cite{TW}) a subbase for the
topology on $M$, one can construct a net indexed by all open subsets
of $M$ in an natural manner from a net indexed by $\Ds$, or even $\Ws$,
alone.

     Tomita--Takesaki modular theory is used to prove the following
results.

\begin{prop}
   (i) \cite{Dri,Lon} Under the above conditions, $\As(W)$ is a
type III$_1$ factor, for every wedge $W \in \Ws$.

   (ii) \cite{Fre,BV} Under the above conditions and with a mild
additional assumption (existence of a scaling limit), also the double
cone algebras, $\As(\Os)$, $\Os \in \Ds$, are type III$_1$ (though not
necessarily factors).
\end{prop}

     The fact that local algebras in relativistic quantum field theory
are typically type III algebras is neither restricted to vacuum
representations nor even to Minkowski space theories. Moreover,
although in a vacuum representation the quasilocal von Neumann algebra
$\As''$ is type I, as seen in Section \ref{states} $\As''$ will not be
type I in GNS representations corresponding to temperature
equilibrium states.  Indeed, there even exists a class of physically
relevant representations in which {\it any} properly infinite
hyperfinite factor can be realized as $\As''$ \cite{DFG,BD}. It is
therefore evident that not only are type I algebras insufficient for
the description of models in quantum statistical mechanics, they are
even less suitable for relativistic quantum field theory.  That is not
to say that they are irrelevant in quantum field theory, for they play
important auxiliary roles, which cannot be discussed here (but see
\cite{BDF,BY,DL,DoL,Haa,Sum82,Sum90,Yng}).

\section{Some Differences of Note}  \label{differences}

     Some of the notable differences between the structure and
predictions of classical (abelian), nonabelian type I, and non-type--I
probability theories will be adumbrated in this section.

\subsection{Entanglement and Bell's Inequalities} \label{Bell}

     Consider a composite system consisting of two subsystems whose
observables are given by the self-adjoint elements of
the von Neumann algebras $\Ms, \Ns \subset \Bs(\Hs)$, respectively.
If these two subsystems are in a certain sense independent, then the
algebras
mutually commute, \ie $\Ms \subset \Ns{}'$. The algebra of observables
of the composite system would be $\Ms \bigvee \Ns = (\Ms \cup \Ns)''$.
A state $\phi$ on $\Ms \bigvee \Ns$ is a {\it product state} if
\begin{equation}  \label{independent}
\phi(MN) = \phi(M) \, \phi(N) \, , \, M \in \Ms \, ,  \, N \in \Ns \, .
\end{equation}
In classical probability, where $\Ms = L^\infty(X_1,\Ss_1,p_1)$ and
$\Ns = L^\infty(X_2,\Ss_2,p_2)$, if a state $\phi$ induced by a
probability measure $p$ on $X_1 \times X_2$ satisfied (\ref{independent}),
one would say that the random variables of the two subsystems are mutually
independent (with respect to $p$). Recall that any (normal) state on
$\Ms \bigvee \Ns$ can be nonuniquely extended to a (normal) state
on $\Bs(\Hs)$, so it is often convenient to view $\phi$ in
(\ref{independent}) as a state on $\Bs(\Hs)$.

     In many applications of quantum theory (though certainly not
all\footnote{See Section \ref{conditional} for a brief discussion of
when this assumption is justifiable.}), the
algebra of observers of the composite system can be taken to be
the tensor product
$\Ms \otimes \Ns \subset \Bs(\Hs) \otimes \Bs(\Hs) \simeq
\Bs(\Hs\otimes\Hs)$,
where $\Ms$ is identified with $\Ms \otimes I$ and $\Ns$ with
$I \otimes \Ns$. A normal state $\phi$ on
$\Ms \otimes \Ns$ is {\it separable}\footnote{also termed decomposable,
classically correlated, or unentangled by various authors} if it is in
the norm closure of the convex hull of the normal product states
on $\Ms \otimes \Ns$, \ie it is a mixture of normal product states.
Otherwise, $\phi$ is said to be {\it entangled}. From the point of view
of what is now called quantum information theory, the primary
difference between classical and noncommutative probability theory is
the existence of entangled states in the latter case. In fact, one has
the following result:

\begin{prop} \cite{Rag}
Every state on $\Ms \otimes \Ns$ is separable if and only if
either $\Ms$ or $\Ns$ is abelian.
\end{prop}

     Hence, if both systems are quantum, \ie both algebras are
noncommutative, then there exist entangled states on the composite
system. Although not understood at that time in this manner, some of
the founders of quantum theory realized as early as 1935
\cite{EPR,Schr} that such entangled states were the source of
``paradoxical'' behavior of quantum theory, as viewed from the vantage
point of classical physics. Today, entangled states are viewed as a
resource to be employed to carry out tasks which cannot be done
classically, \ie only with separable states --- cf. \cite{Key,WW}.

     A primary task of any probability theory is to describe and
estimate the strength of observed correlations. In this connection, a
profound glimpse into the differences between classical, nonabelian
type I and type III probability theories is provided by Bell's
inequalities. We shall only discuss those aspects of Bell's
inequalities which are of immediate relevance to our purposes and
refer the reader to \cite{SW87a,WW} for background and further
references.

     The following definition was made in \cite{SW85}.

\begin{definition} Let $\Ms,\Ns \subset \Bs(\Hs)$ be von Neumann algebras
such that $\Ms \subset \Ns{}'$. The maximal Bell correlation of
the pair $(\cM,\cN)$ in the state $\phi \in \Bs(\Hs)^*$ is
$$\beta(\phi,\cM,\cN) \equiv \sup \, \frac{1}{2} \,
\phi(M_1(N_1+N_2)+M_2(N_1-N_2))\, , $$
where the supremum is taken over all self-adjoint
$M_i \in \cM, N_j \in \cN$ with norm less than or equal to 1.
\end{definition}

     As explained in \eg \cite{SW87a}, the CHSH version
of Bell's inequalities can be formulated in algebraic quantum
theory as
\begin{equation}  \label{bell}
\beta(\phi,\cM,\cN) \leq 1 \, .
\end{equation}
This inequality places a bound on the strength of a certain family
of correlations of observables in $\Ms$, $\Ns$ in the state $\phi$.
This bound is satisfied in {\it every} state, if at least one of the
systems is classical.

\begin{prop} \cite{SW87a}
Let $\Ms,\Ns \subset \Bs(\Hs)$ be mutually commuting von Neumann algebras.
If either $\Ms$ or $\Ns$ is abelian, then $\beta(\phi,\Ms,\Ns) = 1$
for all states $\phi \in \Bs(\Hs)^*$.
\end{prop}

     If, on the other hand, both algebras are nonabelian, then there
always exists a state in which the inequality (\ref{bell}) is violated.
Note that it is known \cite{Cir,SW87a} that
$1 \leq \beta(\phi,\cM,\cN) \leq \sqrt{2}$, for all states $\phi$ on
$\Bs(\Hs)$. For this reason, one says that if
$\beta(\phi,\cM,\cN) = \sqrt{2}$, then the pair $(\Ms,\Ns)$ maximally
violates Bell's inequalities in the state $\phi$.

\begin{prop}\cite{Lan}  \label{landau}
If $\Ms,\Ns \subset \Bs(\Hs)$ are nonabelian, mutually commuting von
Neumann algebras satisfying the Schlieder property, \ie $0 = MN$ for
$M \in \Ms$ and $N \in \Ns$ entails either $M=0$ or $N = 0$, then there
exists a normal state $\phi \in \Bs(\Hs)^*$ such that
$\beta(\phi,\cM,\cN) = \sqrt{2}$.
\end{prop}

     Hence, when $\Ms$ and $\Ns$ are nonabelian, there even exists a
normal state in which Bell's inequalities are maximally
violated. However, the genericity of the states in which (\ref{bell})
is violated, as well as the degree to which it is violated, depends on
finer structure properties of the algebras. We shall only mention
enough results of this type to clearly indicate that there is an
important difference between type I and non-type--I behavior. For
further discussion and references concerning the violation of Bell's
inequalities in algebraic quantum theory, see \cite{Sum97,Red,HC}.

\begin{prop} \label{ours}
Let $\Ms,\Ns \subset \Bs(\Hs)$ be mutually commuting von Neumann algebras.

   (i) \cite{Sum90} If the algebras $\Ms,\Ns$ are type I factors (or
are contained in mutually commuting type I factors), then there exist
infinitely many normal states $\phi \in \Bs(\Hs)^*$ such that
$\beta(\phi,\cM,\cN) = 1$.

   (ii) \cite{SW87b} If $\Ns = \Ms'$, then $\beta(\phi,\cM,\cN) = \sqrt{2}$
for every normal state $\phi \in \Bs(\Hs)^*$ if and only if
$\Ms \simeq \Ms \otimes \Rs_1$, where $\Rs_1$ is the (up to isomorphism)
unique hyperfinite type II$_1$ factor.

\end{prop}

     Note that $\Ms \otimes \Rs_1$ is never a type I algebra.
Maximal violation of Bell's inequalities in every normal state can
only occur in the non-type--I case. In \cite{SW88} it is shown under
quite general physical assumptions in relativistic quantum field
theory that there exist local observable algebras $\As(\Os_1)$,
$\As(\Os_2)$ such that $\beta(\phi,\cA(\Os_1),\cA(\Os_2)) = \sqrt{2}$,
for every normal state $\phi$, and that the circumstances described in
Prop. \ref{ours} (ii) actually obtain. We remark further that under
another set of general physical assumptions \cite{BDF}, the local
algebras $\As(\Os)$ appearing in relativistic quantum field theory
are isomorphic to $\Rs \otimes \Zs$, where $\Zs$ is an abelian
von Neumann algebra and $\Rs$ is the (up to isomorphism) unique
hyperfinite type III$_1$ factor. Since $\Rs \simeq \Rs \otimes \Rs_1$,
the relevance of Prop. \ref{ours} (ii) is reinforced.

     To return briefly to the starting point of this section, states
which violate Bell's inequalities are necessarily entangled. The
converse is not true (cf. \cite{WW} for a discussion and
references). In the now quite extensive quantum information theory
literature, there are various attempts to quantify the degree of
entanglement of a given state (cf. \cite{Key}), but all agree that
maximal violation of inequality (\ref{bell}) entails maximal
entanglement. Only in the non-type--I case is it possible for
every normal state to be maximally entangled.

\subsection{Independence}  \label{independence}

     Another standard topic in probability theory is independence,
which has already been briefly touched upon above. This is an
extensively studied subject, and we shall only mention a few
salient points. For further discussion and references, see
\cite{Sum90,Red,Ham}. As is typical of noncommutative generalizations
of abelian concepts, there are many notions which are distinct
for nonabelian algebras but equivalent in the abelian special case.
We shall discuss only three here.

\begin{definition} Let  $\Ms, \Ns \subset \Bs(\Hs)$ be von
Neumann algebras. The pair $(\Ms,\Ns)$ is \C independent if
for every state $\phi_1$ on $\Ms$ and every state $\phi_2$ on
$\Ns$ there exists a state $\phi$ on $\Bs(\Hs)$ such that
$\phi(M) = \phi_1(M)$, for every $M \in \Ms$ and
$\phi(N) = \phi_2(N)$, for every $N \in \Ns$.
\end{definition}

     So $(\Ms,\Ns)$ is \C independent if every pair of states
on the subsystems represented by the algebras $\Ms,\Ns$ has a
joint extension to the composite system --- in operational terms,
this means that no preparation of one subsystem excludes any
preparation of the other. If every pair of {\it normal} states
on the subsystems has a joint extension to a {\it normal}
state on the composite system, then $(\Ms,\Ns)$ is
said to be \W independent. \W independence implies \C independence
\cite{Sum90,FS}. If $\Ms \subset \Ns{}'$, then these notions are
equivalent \cite{FS}, but \W independence is strictly stronger
when the algebras do not mutually commute \cite{Ham02}. If
$\Ms \subset \Ns{}'$ and $(\Ms,\Ns)$ is \C independent, then
the (not necessarily normal) joint extension can be chosen to be a
product state \cite{Roo}:
$$\phi(MN) = \phi_1(M) \, \phi_2(N) \, , \, M \in \Ms \, , \, N \in \Ns
\, . $$
We reemphasize that the independence expressed by a product state
is the most directly analogous to the notion of independence familiar
from classical probability theory.

     Another of the distinctions between type I and non-type--I
probability theory is enunciated in the following theorem.

\begin{prop}
   (1) If $\Ms \subset \Bs(\Hs)$ is a type I factor and $\Ns = \Ms{}'$,
then $(\Ms,\Ns)$ is \W independent, and given
any normal state $\phi_1$ on $\Ms$ and $\phi_2$ on $\Ns$, the normal
joint extension $\phi$ can be chosen to be a product state.

   (2) If $\Ms \subset \Bs(\Hs)$ is a type III (or II) factor and
$\Ns = \Ms{}'$, then $(\Ms,\Ns)$ is \W independent. But given
any normal states $\phi_1$ on $\Ms$ and $\phi_2$ on $\Ns$,
the normal joint extension cannot be chosen to be a product state.
$\phi_1$ and $\phi_2$ have a joint extension to a product state
$\phi$, but $\phi$ cannot be normal. Hence, $\phi$ is only
finitely additive on $\PH$ and not $\sigma$-additive.

\end{prop}

\begin{proof}
(1) is well known, and (2) is discussed in
\cite{Sum90,FS} for the type III case, but for the reader's benefit
a proof will be sketched here. Since $\Ms$ and $\Ns$ are
commuting factors, they satisfy the Schlieder property and thus are \C
independent \cite{Roo}.  This entails that the states $\phi_1$ and
$\phi_2$ have a joint extension to a product state on
$\Ms \bigvee \Ns$ \cite{Roo} and that $(\Ms,\Ns)$ is
\W independent \cite{FS}.  If there did exist a normal product
state across $(\Ms,\Ns)$, then \cite{Tak58} $\Ms \bigvee \Ns$ is
isomorphic to $\Ms \otimes \Ns$, so that there exists a type I factor
$\Ls$ such that \cite[Th. 1 and Cor. 1]{DL} $\Ms \subset \Ls \subset
\Ns' = \Ms$, \ie $\Ms = \Ls$, a contradiction unless $\Ms$ is type
I. If $\Ms$ is type I, then $\Ms \bigvee \Ns \simeq \Ms \otimes \Ns$
(see \eg \cite{DL}), and the state $\phi_1 \times \phi_2$ on
$\Ms \otimes \Ns$ (defined by
$(\phi_1 \times \phi_2)(\sum_i M_i N_i) = \sum_i \phi_1(M_i) \,
\phi_2(N_i)$)
precomposed with the isomorphism implementing
$\Ms \bigvee \Ns \rightarrow \Ms \otimes \Ns$ is a normal joint
extension of $\phi_1$ and $\phi_2$ on $\Ms \bigvee \Ns$.
\end{proof}

     Hence, there are many normal product states on $(\Ms,\Ms{}')$ in
the type I case, but there are none when $\Ms$ is type II or III.
It is perhaps worthwhile in this connection to mention that
in a physical setting with a specified dynamics implemented unitarily
as in (\ref{ground}), long experience has indicated that the following
folklore is correct: it takes an infinite amount of energy to create a
nonnormal state from a normal state. Thus, the nonnormal product state
in the type III case is not likely to be physically realizable,
whereas the normal product state in the type I case is prepared in
actual laboratories every day. This supports the physical relevance of
the distinction drawn in the previous theorem.

     A further notion of statistical independence was proposed in
\cite{Lic,Kra}.

\begin{definition}
A pair $(\Ms,\Ns)$ of von Neumann algebras on $\Hs$ is strictly
local (in the sense of vector states) if for any
$P \in \PM$ and any unit vector $\Phi \in \Hs$ there
exists a unit vector $\Psi \in P\Hs$ such
that $\langle \Phi ,N\Phi \rangle = \langle \Psi ,N\Psi \rangle$
for all $N \in \Ns$.
\end{definition}

     As shown in \cite{Sum90}, this condition implies a more
transparent condition of independence called {\it strict locality}:
for every $P \in \PM$ and every normal state $\phi_2$ on $\Ns$, there
exists a normal state $\phi \in \Bs(\Hs)^*$ such that $\phi(P) = 1$
and $\phi(N) = \phi_2(N)$, for all $N \in \Ns$. Hence, no preparation
on the subsystem represented by $\Ns$ can exclude the truth of any
proposition in $\Ms$ (cf. \cite{Sum90} for further discussion of the
relation between these properties).  Both of these properties imply \C
independence \cite{FS}, and \W independence implies strict locality
\cite{Sum90}. The following result thus associates the type
III structure property with an independence property having physical
significance.

\begin{prop} \cite{Lic} Let $\Hs$ have dimension greater than 1,
$\Ms \subset \Bs(\Hs)$ be a von Neumann factor and $\Ns = \Ms{}'$.
The pair $(\Ms,\Ns)$ is strictly local in the sense of vector states
if and only if $\Ms$ is type $III$.
\end{prop}

\subsection{Conditional Expectations} \label{conditional}

     A concept of central importance in probability theory is that of
conditional expectation. We present the classical notion from the
point of view of operator algebra theory in order to motivate the
general definition in the noncommutative setting. Although the concept
extends to certain classes of unbounded random variables and
corresponding classes of unbounded operators, we shall only discuss
the bounded case here.

     Let $(X,\Ss,p)$ be a probability space and $\Ts$ be a
sub-$\sigma$-algebra of $\Ss$. Then for any $f \in L^\infty(X,\Ss,p)$
there exists a unique $E_\Ts(f) \in L^\infty(X,\Ts,p)$ such that
\begin{equation} \label{conditionaleq}
\int \, \chi_T f \, dp = \int \, \chi_T E_\Ts(f) \, dp \, , \, T \in \Ts \,
.
\end{equation}
$E_\Ts(f)$ is called the {\it conditional expectation} of $f$ with respect
to $\Ts$. Its existence is assured by the Radon--Nikodym Theorem.
This then yields a linear map
$E_\Ts : L^\infty(X,\Ss,p) \rightarrow L^\infty(X,\Ts,p)$ with the
following properties: $E_\Ts(f) = f$, for all $f \in L^\infty(X,\Ts,p)$
and $\Vert E_\Ts(f) \Vert \leq \Vert f \Vert$, for all
$f \in L^\infty(X,\Ss,p)$. So
$E_\Ts : L^\infty(X,\Ss,p) \rightarrow L^\infty(X,\Ts,p)$ is a
projection of norm 1. Moreover,
$E_\Ts(\sup_\alpha f_\alpha) = \sup_\alpha E_\Ts(f_\alpha)$,
for any uniformly bounded increasing net $\{ f_\alpha \}$ of positive
elements of $L^\infty(X,\Ss,p)$.

     The noncommutative generalization can therefore be formulated
as follows. Let $\Ms \subset \Bs(\Hs)$ be a von Neumann algebra
and $\Ns \subset \Ms$ be a subalgebra. A {\it conditional expectation}
(on $\Ms$ relative to $\Ns$) is a linear map $E_\Ns : \Ms \rightarrow \Ns$
with norm 1 whose restriction to $\Ns$ is the identity. If, in addition,
$E_\Ns$ satisfies the continuity condition
$E_\Ns(\sup_\alpha A_\alpha) = \sup_\alpha E_\Ns(A_\alpha)$,
for any uniformly bounded increasing net $\{ A_\alpha \}$ of positive
elements of $\Ms$, $E_\Ns$ is a {\it normal conditional expectation}.
Given $M \in \Ms$, $E_\Ns(M)$ is called the conditional expectation
of $M$ with respect to $\Ns$.

     Of course, this definition does not yet reproduce the essential
(for probability theory) condition (\ref{conditionaleq}), whose
generalization requires specifying a normal state $\phi$ on $\Ms$:
\begin{equation} \label{conditionalnon}
\phi(PM) = \phi(PE_\Ns(M)) \, , \, M \in \Ms \, , \, P \in \Ps(\Ns) \, .
\end{equation}
However, Tomiyama proved the following result.

\begin{prop} \cite{Tom} \label{Tomiyama}
If $E_\Ns : \Ms \rightarrow \Ns$ is a conditional expectation, then
for any $0 \leq M \in \Ms$ one has $E_\Ns(M) \geq 0$. Moreover,
$E_\Ns(NM) = NE_\Ns(M)$, for all $N \in \Ns$, $M \in \Ms$. Consequently,
$E_\Ns(M)^* = E_\Ns(M^*)$ and $E_\Ns(MN) = E_\Ns(M)N$, for all
$N \in \Ns$, $M \in \Ms$.
\end{prop}
One therefore sees that condition (\ref{conditionalnon}) is
equivalent to $\phi = \phi \circ E_\Ns$. One says that a conditional
expectation $E_\Ns : \Ms \rightarrow \Ns$ is {\it faithful} if
$0 \leq M \in \Ms$ and $E_\Ns(M) = 0$ entail $M = 0$. In the
classical case, the conditional expectations {\it are} faithful.
Thus, given a normal state $\phi$ on $\Ms$, a faithful normal
conditional expectation $E_\Ns : \Ms \rightarrow \Ns$ such that
$\phi \circ E_\Ns = \phi$ is the proper generalization of the classical
concept.

     As indicated above, in the abelian case, for any subalgebra $\Ns
\subset \Ms$ and any normal state $\phi$ on $\Ms$ there exists a
unique faithful normal conditional expectation on $\Ms$ relative to
$\Ns$ leaving $\phi$ invariant. The same is not true in general. Once
again, we can only discuss certain aspects of the matter.

     First, let us appreciate the significance of the normality of the
conditional expectation in the general case. A factor
$\Ms \subset \Bs(\Hs)$ is said to be
{\it injective} if there exists a (not necessarily normal) conditional
expectation $E : \Bs(\Hs) \rightarrow \Ms$. It is certainly not the
case that every factor is injective, but type I factors are injective,
as are the non-type--I factors which typically arise in quantum
statistical mechanics \cite{AWo,AWy} and in quantum field theory
\cite{BDF}, since hyperfinite factors are injective.  Requiring that a
conditional expectation be normal imposes serious constraints on the
strucure, as the following result makes clear.

\begin{prop} \cite{Tom70}  \label{structurethm}
Let $\Ms, \Ns \subset \Bs(\Hs)$ and $\Ns$ be a subalgebra of $\Ms$.

   (i) Let $\Ms$ be a factor and $\Ns$ be type III. If there exists a
normal conditional expectation $E_\Ns : \Ms \rightarrow \Ns$, then
$\Ms$ is type III.

   (ii) Let $\Ms$ be semifinite, \ie have no direct summand of type III.
If there exists a normal conditional expectation
$E_\Ns : \Ms \rightarrow \Ns$, then $\Ns$ is semifinite.

   (iii) If $\Ms$ is type I and there exists a normal conditional
expectation $E_\Ns : \Ms \rightarrow \Ns$, then $\Ns$ is type I.

\end{prop}

     Adding the condition that the normal conditional expectation
leaves a distinguished state invariant is even more restrictive.
Takesaki has given a characterization of this situation when the
state is faithful, but to
state the result properly, a few preparations must be made.
Given a faithful normal state $\phi$ on $\Ms$, there
is uniquely associated a one-parameter group of automorphisms,
$\sigma_t$, $t \in \RR$, of $\Ms$ called the {\it modular automorphism
group} corresponding to $\phi$, and $(\Ms,\{\sigma_t\}_{t \in \RR})$
forms a (\W)dynamical system for which $\phi$ is a KMS state at
inverse temperature $\beta = 1$ \cite{Takm,BRII,KR,Tak}.

\begin{prop} \cite{Tak92}
Let $\phi$ be a faithful normal state on the von Neumann algebra
$\Ms \subset \Bs(\Hs)$ and let $\Ns \subset \Ms$ be a subalgebra.
There exists a faithful normal conditional expectation
$E_\Ns : \Ms \rightarrow \Ns$ such that $\phi \circ E_\Ns = \phi$
if and only if $\sigma_t(\Ns) \subset \Ns$, for all $t \in \RR$.
Under these circumstances, if $\Ns^c = \Ns' \cap \Ms$ is the
{\it relative commutant} of $\Ns$ in $\Ms$ and $\Ns$ is a factor,
then $\Ns \bigvee \Ns^c \simeq \Ns \otimes \Ns^c$ and $\phi$ is a
product state on $\Ns \bigvee \Ns^c$. Moreover, $E_\Ns$ is unique.
\end{prop}

\noindent It should now be clear that the existence of such a
conditional expectation in truly noncommutative probability theory is
more the exception than the rule. Nonetheless, they do exist in
physically interesting circumstances (cf. \cite{DoL,BDF,Par,CO}) and
are quite useful.

     As mentioned in Section \ref{Bell}, it is of physical relevance
to know under which conditions the algebra of observables of the
composite system $\Ms \bigvee \Ns$ is isomorphic to $\Ms \otimes \Ns$,
and the matter has received a lot of attention from mathematical
physicists and operator algebraists. It is fitting to give here a
characterization in terms of the existence of a normal
conditional expectation.\footnote{For a generalization to the nonfactor
case, see \cite{Tom70}.}

\begin{prop} \cite{Tak58}
Let $\Ms, \Ns \subset \Bs(\Hs)$ be mutually commuting factors. Then
$\Ms \bigvee \Ns$ is isomorphic to $\Ms \otimes \Ns$ if and only if
there exists a normal conditional expectation
$E : \Ms \bigvee \Ns \rightarrow \Ms$.
\end{prop}

\noindent For operationally motivated sufficient conditions entailing
$\Ms \bigvee \Ns \simeq \Ms \otimes \Ns$, see \cite{BS}.

    It may be instructive to see simple examples of such conditional
expectations. Let $A = A^* \in \Bs(\Hs)$ have purely discrete spectrum
consisting of simple eigenvalues and $P_i$, $i \in \IN$, denote the
projections onto the corresponding one dimensional eigenspaces. In
quantum measurement theory, the following map is associated with the
so--called {\it projection postulate} (cf. \cite{Neumann1932}):
$$T_A(B) = \sum_i P_i B P_i \, , \, B \in \Bs(\Hs) \, . $$
Note that $AT_A(B) = T_A(B)A$, for all $B \in \Bs(\Hs)$, and that
for any $B \in \{ A \}'$, $T_A(B) = B$ (see Prop. \ref{functionalcalculus}).
Hence, $T_A : \Bs(\Hs) \rightarrow \{ A \}'$ is a normal conditional
expectation of a type I algebra onto a type I algebra which preserves
states with density matrix $P_i$.

     A situation commonly arising in nonrelativistic quantum
theory (and quantum information theory) is a composite system
consisting of two subsystems, which is modelled by
$\Bs(\Hs_1) \otimes \Bs(\Hs_2)$, prepared in a state $\phi$ determined
by the density matrix $\rho$ on $\Hs_1 \otimes \Hs_2$ given by
$\rho = \rho_1 \otimes \rho_2$, where $\rho_i$ is a density matrix
on $\Hs_i$, $i = 1,2$. Let $\tr_1$ represent the trace on $\Bs(\Hs_1)$.
Then the map
$E_2 : \Bs(\Hs_1) \otimes \Bs(\Hs_2) \rightarrow I_1 \otimes \Bs(\Hs_2)$
determined on special elements of the form $A_1 \otimes A_2$,
$A_1 \in \Bs(\Hs_1)$, $A_2 \in \Bs(\Hs_2)$, by
$$E_2(A_1 \otimes A_2) = (\tr_1(\rho_1 A_1) \, I_1) \otimes A_2 =
\tr_1(\rho_1 A_1) (I_1 \otimes A_2)$$
and extended to $\Bs(\Hs_1) \otimes \Bs(\Hs_2)$ by linearity and
continuity, is a faithful normal conditional expectation. Moreover,
it leaves the state $\phi$ invariant, since
\begin{eqnarray*}
\phi(E_2(A_1 \otimes A_2))&  = & \phi(\tr_1(\rho_1 A_1) (I_1 \otimes A_2))
= \tr_1(\rho_1 A_1) \cdot \phi(I_1 \otimes A_2) \\
& = & \tr_1(\rho_1 A_1) \cdot \tr_1(\rho_1 I_1) \cdot \tr_2(\rho_2 A_2) =
\tr_1(\rho_1 A_1) \cdot \tr_2(\rho_2 A_2) \\
& = & \phi(A_1 \otimes A_2) \, .
\end{eqnarray*}
This is a map from a type I algebra to a type I algebra.

     A related, but less elementary example is provided by the one
dimensional lattice gas discussed above. Referring to the tracial state
$\tau$ on $\As''$  constructed in Section \ref{states}, let
$\Lambda \subset \Zs$ be finite and consider the algebra
$\As(\Lambda)$ defined in Section \ref{structure}, which
is the subalgebra of $\As$ generated by the observables in $\Lambda$.
Then define the map
$E_\Lambda : \As'' \rightarrow \As(\Lambda)''$
by considering an arbitrary element
$\otimes_{i \in \Lambda_0} A_i \in \As_0$ (for some finite
$\Lambda_0 \subset \Zs$), setting
$$E_\Lambda(\otimes_{i \in \Lambda_0} A_i) =
\otimes_{i \in \Lambda_0} \widetilde{A_i} \, , $$
where $\widetilde{A_i} = \widehat{\tr}_i(A_i) I_i$, if $i \notin \Lambda$,
and $\widetilde{A_i} = A_i$, if $i \in \Lambda \cap \Lambda_0$, and
extending by linearity and continuity. $E_\Lambda$ is then a faithful
normal conditional expectation leaving the state $\tau$ invariant. This
is a map from a type II algebra to a type I algebra.

\subsection{Further Comments on Type I versus Type III} \label{contrast}

     In this section, we briefly discuss some further distinctions
between the type I and type III cases, which, although perhaps
not strictly probabilistic in nature, are of direct physical
relevance.

     As previously mentioned, every nonzero projection $P$ in a type
III algebra $\Ms \subset \Hs$ is infinite. This entails that $P\Hs$ is
an infinite dimensional subspace of $\Hs$. So, in particular, if
$A = A^* \in \Ms$ and $P$ is the projection onto an eigenspace of $A$,
then since $P \in \Ms$, it follows that every eigenvalue of $A$ is
infinitely degenerate. Thus, the many papers in the physics literature
which restrict their considerations to observables with simple
eigenvalues tacitly exclude all the physical situations in which
type III algebras arise.

     Typically in relativistic quantum field theory, both the algebra
of observables localized in a region $\Os$ and that localized in its
causal complement $\Os'$ are type III, and they are each other's
commutant: $\As(\Os)' = \As(\Os')$. In such a circumstance,
given any orthogonal projection
$P \in \As(\Os)$ and any state $\phi$ on the quasilocal algebra $\As$,
it is possible to change $\phi$ into an eigenstate of $P$ with an
operation strictly localized in $\Os$, \ie an operation which does not
disturb the expectation of any observable localized in $\Os'$
\cite{Lic}. Namely, since $P$ is equivalent to $I$ in $\As(\Os)$,
there exists a partial isometry $W \in \As(\Os)$ such that $P = WW^*$
and $I = W^*W$. The state $\phi_W(\cdot) = \phi(W^* \cdot W)$ then
satisfies
$$\phi_W(P) = \phi(W^* P W) = \phi(W^* W W^* W) = \phi(I) = 1 \, , $$
on the one hand, and, for all $A \in \As(\Os') = \As(\Os)'$,
$$\phi_W(A) = \phi(W^* A W) = \phi(A W^* W) = \phi(A) \, , $$
on the other. Moreover, since the algebra $\As(\Os)$ is usually a
type III$_1$ factor, the transitivity of the action of
the group of unitaries on the normal state space of such a factor
\cite{CS} entails that given any two normal states $\phi,\omega$
on $\As(\Os)$ and an $\epsilon > 0$, there exists a unitary
$W \in \As(\Os)$ such that
$$\vert \omega(A) - \phi_W(A) \vert \leq \epsilon \Vert A \Vert \, ,$$
for all $A \in \As(\Os)$. In other words, every normal state can be
prepared locally with arbitrary precision from any other normal
state. These facts rely upon properties of type III algebras which
do not obtain for type I algebras.

     These and other distinctions are sources of errors found in
the literature which essentially amount to applying reasoning valid
for type I quantum theory to type III quantum theory. A notable
example of this is the argument given in \cite{Heg}, which purports to
demonstrate that relativistic quantum field theory violates
causality. In an immediate retort \cite{BY} (see \cite{Yng} for a
possibly more accessible explanation), it was pointed out that
the argument employed in \cite{Heg} rested upon an inadmissible use of
type I reasoning.

\section{Closing Words}  \label{closing}

     So, what is ``quantum probability theory''? Some authors use the
term synonymously with noncommutative probability theory.  And others
use it to mean noncommutative probability theory with (some of) the
additional structures which physical considerations add to basic von Neumann
algebra theory. We regard this as a matter of personal taste. In
either case, we have endeavored to make clear at least two main
points: (1) both classical probability theory and quantum theory are
special cases of noncommutative probability theory, and (2) there are
significant differences between the type I and non-type--I quantum
theories. The former point emphasizes the existence of an elegant, unifying
framework within which the latter can be studied and better
understood. These probability theories form a spectrum with the
abelian case located at one extreme, the type III case at the other,
and the standard type I quantum theory located squarely between them.

\end{document}